\documentclass[11pt,a4paper]{jpconf}

\usepackage[usenames,dvipsnames]{color}
\usepackage{amsmath}
\usepackage[final]{pdfpages}
\usepackage{graphicx}
\usepackage{geometry}
\begin{document}

\title{Coarsening and percolation in the Ising Model with quenched disorder}
\author{F Insalata$^{1,3}$, F Corberi$^2$,
 L F Cugliandolo$^3$ and M Picco$^3$}
\address{ $^1$ Dipartimento di Fisica ``E.~R. Caianiello'', Universit\`a  di Salerno, 
via Giovanni Paolo II 132, 84084 Fisciano (SA), Italy.}

\address{$^2$ INFN, 
Gruppo Collegato di Salerno, and CNISM, Unit\`a di Salerno, Universit\`a  di Salerno, 
via Giovanni Paolo II 132, 84084 Fisciano (SA), Italy.}

\address{$^3$ Sorbonne Universit\'es, 
Universit\'e Pierre et Marie Curie - Paris 6, Laboratoire de Physique
Th\'eorique et Hautes Energies, 
4, Place Jussieu, Tour 13, 5\`eme \'etage,
75252 Paris Cedex 05, France.}
\ead{ferdinsa@live.it}

\begin{abstract}
Through large-scale numerical simulations, we study the phase ordering kinetics of the $2d$  Ising Model after a zero-temperature quench from 
a high-temperature
 homogeneous initial condition. 
Analysing the behaviour of two important quantities  -- the winding angle and  the pair-connectedness -- we reveal the presence 
of a percolating structure in the pattern of domains.  We focus on the pure case 
and on the random field and random bond Ising Model. 
\end{abstract}
\section{Introduction}
\label{intro}

Phase ordering kinetics,  the  ordering of a system via domain growth after a quench from
the homogeneous phase into one with broken symmetry, has attracted 
great interest in the last 50 years \cite{Bray94}. 

A simple example is a ferromagnet, instantaneously cooled (quenched)
from above to below the critical point. The initial equilibrium state becomes unstable after
the quench and evolves towards one of the two possible symmetry-related
ordered configurations with opposite magnetizations. Relaxation toward the new equilibrium state 
occurs slowly (i.e. not exponentially fast) by the formation and growth (coarsening) of domains  of the two equilibrium phases (i.e. group of aligned spins).
This domain growth is driven by superficial tension, i.e. the interfaces tend to become flatter  due to energetic reasons. Over time, the smallest domains disappear so that 
the typical size $R(t)$ of the remaining ones increases.

One theoretical approach to this class of problem
is the kinetic Ising Model (IM), originally introduced by Glauber \cite{Glauber63}. 
One starts with a Ising system in which each spins is randomly oriented, a situation that can be described as 
equilibrium at infinite temperature, $T \rightarrow \infty$.
Then, the evolution of the system is treated as a Markov chain, with appropriate transition probabilities.
The dynamical evolution can  be simulated 
through standard Monte Carlo methods \cite{Cugliandolo14}.
In the thermodynamic limit, the coarsening process goes on indefinitely, with none of the two equilibrium phases (up and down spins in this case) prevailing 
at any given finite time. This means that the magnetic system does not develop a magnetization, or equivalently that on average up spins are in the same number as down spins.

It was recently shown \cite{Arenzon07,Sicilia07,Blanchard14} that, after a transient from the quench, some of the morphological features 
of the coarsening domains in an ordinary, homogeneous IM are the same as those of a geometrical lattice at the percolation threshold $p_c$ of random site percolation
 \cite {Christensen}.
 In such condition, a spanning (percolating) cluster is present.
 In the coarsening IM, we observe a percolating
domain of either up or down spins.
This fact may seem paradoxical, since percolation is a non-interacting problem, whereas coarsening
is driven by spin-spin interactions. Moreover, the absence of magnetization in the coarsening
system means that the percolative structures are present when $1/2$ of spins are up (down). For a $2d$ square lattice,
the same geometry we will consider for the IM, the critical percolation threshold is $p_c \simeq 0.59$, 
when occupying sites randomly. 
This means that we observe percolation in the corsening IM when the occupation probability of up (down) spins is $1/2 <p_c$. 
This is not a contradiction, however, since after the transient $t_p$, the spin configuration 
is no longer random, having been shaped by the spin-spin interactions. 
It is, indeed, only after the coarsening dynamics has gone on for a transient that percolative 
features appear, while they are absent right after the quench,
a sign that they are due to interactions. 
This can be phrased by saying that the spin-spin interactions lower the percolation threshold from $p_c  \simeq 0.59$ to $1/2$. 
Moroever, although random uncorrelated percolation and correlated coarsening coexist on the same lattice, 
they live on separate length scales : 
the percolative properties are not present, at any given time, on length scales that are smaller than $R(t)$, 
the typical size of the domains,  i.e. the correlated regions. 

In a recent work \cite{Corberi17}, we showed that this phenomenology is also found in two types of inhomogeneous IM, where structural (quenched) disorder is present. In particular, we focussed on the random field Ising Model (RFIM) and random bond Ising Model (RBIM).

In this paper, after a brief review of the models considered and of the Monte Carlo algorithm implemented (Sec. \ref{models}), 
we define the relevant observables in Sec. \ref{quantities} and present the basic results which highlight the occurrence of the percolation transition in Sec. \ref{results}.

\section{Models and algorithm}
\label{models}
We consider a $2d$ square lattice of linear size $L$, and the Hamiltonian of our models is 

\begin{equation}
{\cal H}(\{S_i\})=-\sum _{\langle ij\rangle}J_{ij}S_iS_j+\sum _i H_iS_i,
\label{genham}
\end{equation}
where $S_{i}$ are the Ising variables and $\langle ij\rangle$ are two nearest-neighboring sites. The different models considered in our study are determined by the
properties of the coupling constant $J_{ij}$ and of the external field $H_i$, as follows.

{\bf Homogeneous IM:} In this case, $H_i\equiv 0$ and $J_{ij}\equiv J_0$, i.e. we recover the ordinary IM. We also refer to it as the clean or pure model.

{\bf RFIM:} The couplings are constant 
 $J_{ij}\equiv J_0$, while the external fields $H_i$ are
drawn from a symmetric bimodal distribution and uncorrelated in space, i.e. $H_i = \pm h$, each with probability one half.
We refer to this model as the RFIM.

{\bf RBIM: }  No external fied, $H_i\equiv 0$,  while the couplings 
are $J_{ij}=J_0+\delta _{ij}$,
where $\delta _{ij}$ are independent random numbers drawn from a uniform distribution in 
$[-\delta,+\delta]$. We keep  $\delta <J_0$ in order to avoid frustration effects. We refer to this model as the RBIM.

In the case of the RBIM, coarsening via domain growth still takes place in the absence of frustration, which is the case we have considered. In the case of the RFIM, 
the Imry-Ma argument \cite{Imry75} tells us that ordering takes place up to a length 
\begin{equation}
\ell _{IM}\sim \left (\frac{h}{J_0}\right )^{\frac{2}{d-2}}.
\label{im}
\end{equation}
As we will see below, we focus on the case in which $\ell _{IM} \rightarrow \infty$, for which the system orders in any dimension.

As mentioned above, we can treat the dynamical evolution of the system as a Markov chain and simulate it through the Metropolis algorithm. 
Specifically, our algorithm flips single spins with Glauber transition rate 
\begin{equation}
w(S_i\to -S_i)= \frac{1}{2}\left [1-S_i\tanh \left (\frac{H^W_i+H_i}{T}\right )\right ],
\label{trate}
\end{equation}
where the Weiss Field $H^W_i$, is given by 
\begin{equation}
H^W_i=\sum _{j\in  nn(i)}J_{ij}S_j,
\end{equation}
with $j$ running over the nearest-neighbors of the spin $S_i$.

For our simulations, the protocol consists in preparing a system of uncorrelated spins with no magnetization -- equilibrium at $T \rightarrow \infty$ -- and letting it evolve through single-spin flip according to Eq. (\ref{trate}). Performing the quench to $T_f$ amounts to inserting the value of $T_f$ into Eq. (\ref{trate}). We have let
 $T_f \rightarrow 0$ keeping the ratios  $\epsilon =\frac{\delta}{T_f}$ (for  RBIM) and $\epsilon =\frac{h}{T_f}$ (for RFIM) finite.  

First, let us notice that this limit entails $\ell _{IM} \rightarrow \infty$, as mentioned above. Moreover, this simplifies the form of the transition rates, which depend only 
on the parameter $\epsilon$. Not only we obtain a theory that depends on a single parameter, but simulations are sped up and thermal noise is reduced. More details on 
this algorithm and on its reliability can be found in \cite{Burioni07, Corberi01}.

\section{Key observables}
\label{quantities}

The first quantity we consider, an essential one for the study of the phase ordering kinetics of a system, is the typical domain size $R(t)$. 
This is also referred to as \textit{growing length}, since it grows while the system orders through domain growth.
We have computed it as the inverse of the density  of defects, i.e. dividing the number of couple of unaligned spins by $L^2$. 
As discussed in \cite{Puri09}, this is a standard method to estimate $R(t)$.
We have then performed, as for all other quantities, an ensemble average, typically on 
$10^5 - 10^6$ realizations. $R(t)$ grows as $t^{1/2}$ in the pure system and slower and slower for increasing values of the parameter $\epsilon$ controlling the strength of disorder. More details can be found in \cite{Corberi15c}.

In order to probe the emergence of  percolation in the coarsening process, we used two observables : the average squared winding angle and the pair-connectedness function.
They are naturally defined on a geometrical lattice in which a fraction $p$ of the sites is occupied. They can be measured on an Ising lattice by identifying occupied sites as, say, up spins and empty sites as down spins, or vice-versa. A domain can then be considered as a cluster of up (or down) spins and the above quantities can be computed with respect to the domains of the system. 

The (average squared) winding angle is defined considering the external contours -- hulls -- of a cluster of occupied sites. 
One considers two points $P$ and $Q$ at a distance $x$ along the hull, and the tangent to the hull in these two points.
The angle between the two tangents, measured counterclockwise in radiants units, is the winding angle $\theta(P,Q)$.
By fixing $x$, one can calculate the average winding angle for two points separated by this distance, $\langle \theta(x) \rangle$.
It can be exactly proven, through methods of conformal 
field theory, that on a lattice at the percolation threshold  for the  hull of any cluster we have \cite{Duplantier, Wilson}
\begin{equation}
\langle \theta ^ 2(r) \rangle = a + \frac{4k}{8+k} \ln r,
\label{wind_angle}
\end{equation}
where $a$ is a constant and $k=6$. This equation holds when $r \gg r_0$, $r_0$ being the lattice spacing, i.e. when the microscopic structure of the
lattice can be ignored. Therefore, we will report numerical data only for the largest domain of the system, so to have the largest possible
range in which $r\gg r_0$.

The pair-connectedness function is defined in percolation theory as the probability that two occupied sites at distance
$r$ belong to the same cluster.
From $2d$ percolation theory, we know that at $p=p_c$ \cite{Stauffer,Saberi}
\begin{equation}
C_{perc}(r,r_0)\sim \left ( \frac{r}{r_0}\right )^{-2\Delta}
\label{connectperc}
\end{equation}
where $r_0$ is the unit of length, usually the lattice spacing.
Eq. ~(\ref{connectperc}) holds for $r \gg r_0$, with the critical exponent $\Delta =5/48$. 
In the coarsening system we measure the pair-connectedness  at any given time, identifying all the
domains of up and down spins and then computing
\begin{equation}
C(r,t)=\frac{1}{4 L^2}\sum _i \sum _{i_r} \langle \delta _{S_i,S_{i_r}} \rangle,
\label{connect} 
\end{equation}
where $\delta _{S_i,S_j}=1$ if the two spins belong to the same domain -- namely they are
aligned and there is a path of aligned spins connecting them -- and $\delta _{S_i,S_j}=0$
otherwise.

\section{Numerical results}
\label{results}
We now show how measuring the winding angle and the pair-connectedness on the domains of the coarsening IM highlights that the morphological properties of the 
domain patchwork are those of critical percolation. 
\begin{figure}[h!]
\centering
\rotatebox{0}{\resizebox{0.55\textwidth}{!}{\includegraphics{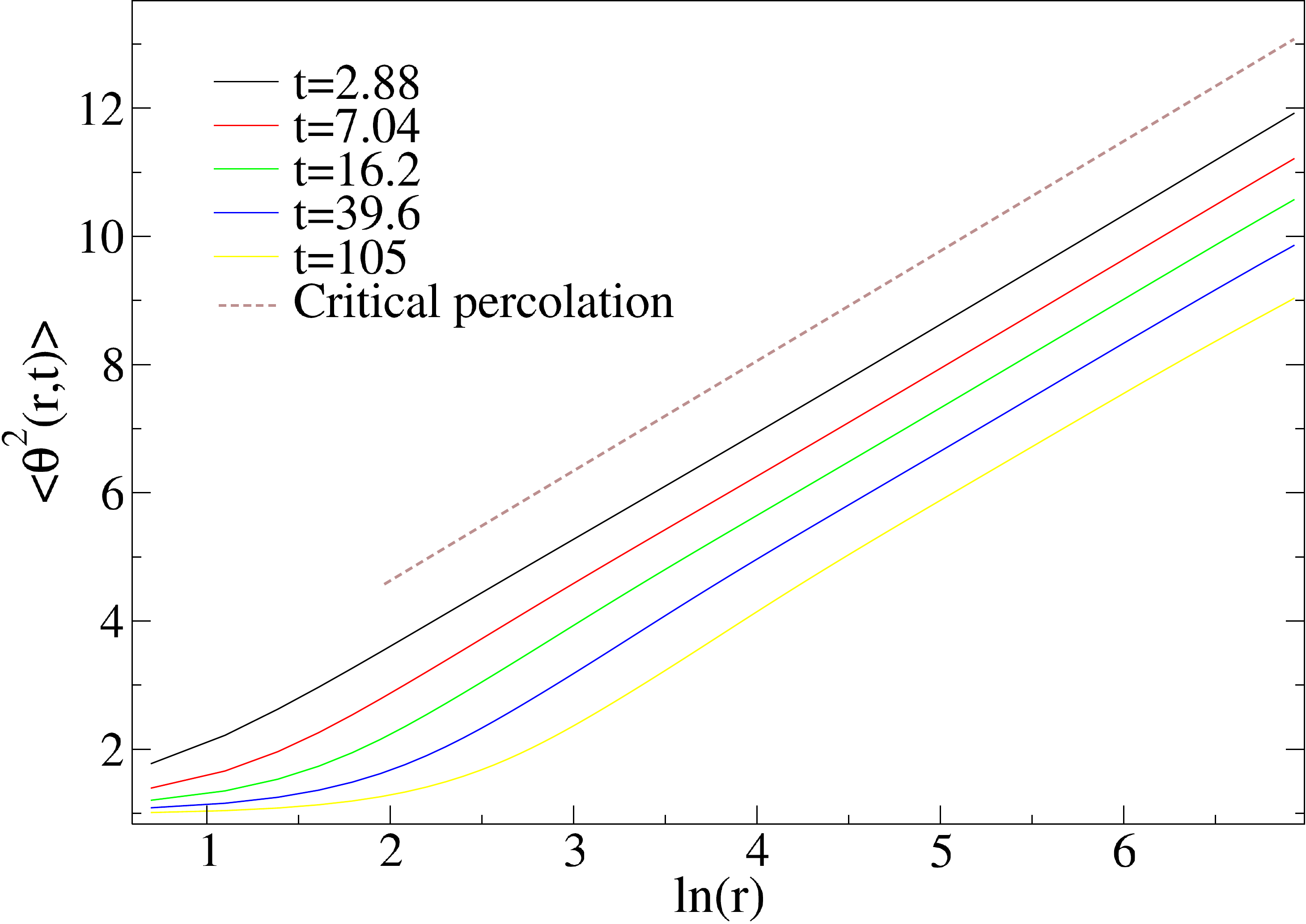}}}
\caption{Average squared winding angle for the domains of the largest cluster of the pure $2d$ IM, plotted against
 $\ln r$, the distance $r$ along the hull. 
Results are relative to a lattice with $L=640$. In the key we report the measurement times. The dashed line is 
Eq.  (\ref{wind_angle}) with $k=6$, the value of critical percolation.}
\label{winding_pure_basic}
\end{figure}

In Fig. \ref{winding_pure_basic} we report results for the clean model of size $L=640$, for various measurement times.
The dashed blue line is the analytical curve, namely Eq.  (\ref{wind_angle}) with $k=6$.

Apart from an upward vertical shift of the curves for increasing measurement times -- whose interpretation is discussed in \cite{Corberi17} --
for each curve we observe two behaviors : an initial part (short distances)
in which the analytical expression is not followed, and a large-distance regime where  numerical
data are very well described by Eq.  (\ref{wind_angle}). Indeed, we fitted the  slope of 
the purple curve  corresponding to $R(t) = 2.68$ and  $t=3.13$ in the range $4<\ln (x) <7$ ,  and we obtained $k=5.94$, 
in excellent agreement with the value  $k=6$ for critical percolation. 
This signals that over sufficiently large distances
the largest cluster has the morphological properties of percolation.
Interestingly, we see that the crossover between the two behaviors takes place at a distance $l_t$ which increases with the measurement time.

\begin{figure}[h!]
\centering
\rotatebox{0}{\resizebox{0.55\textwidth}{!}{\includegraphics{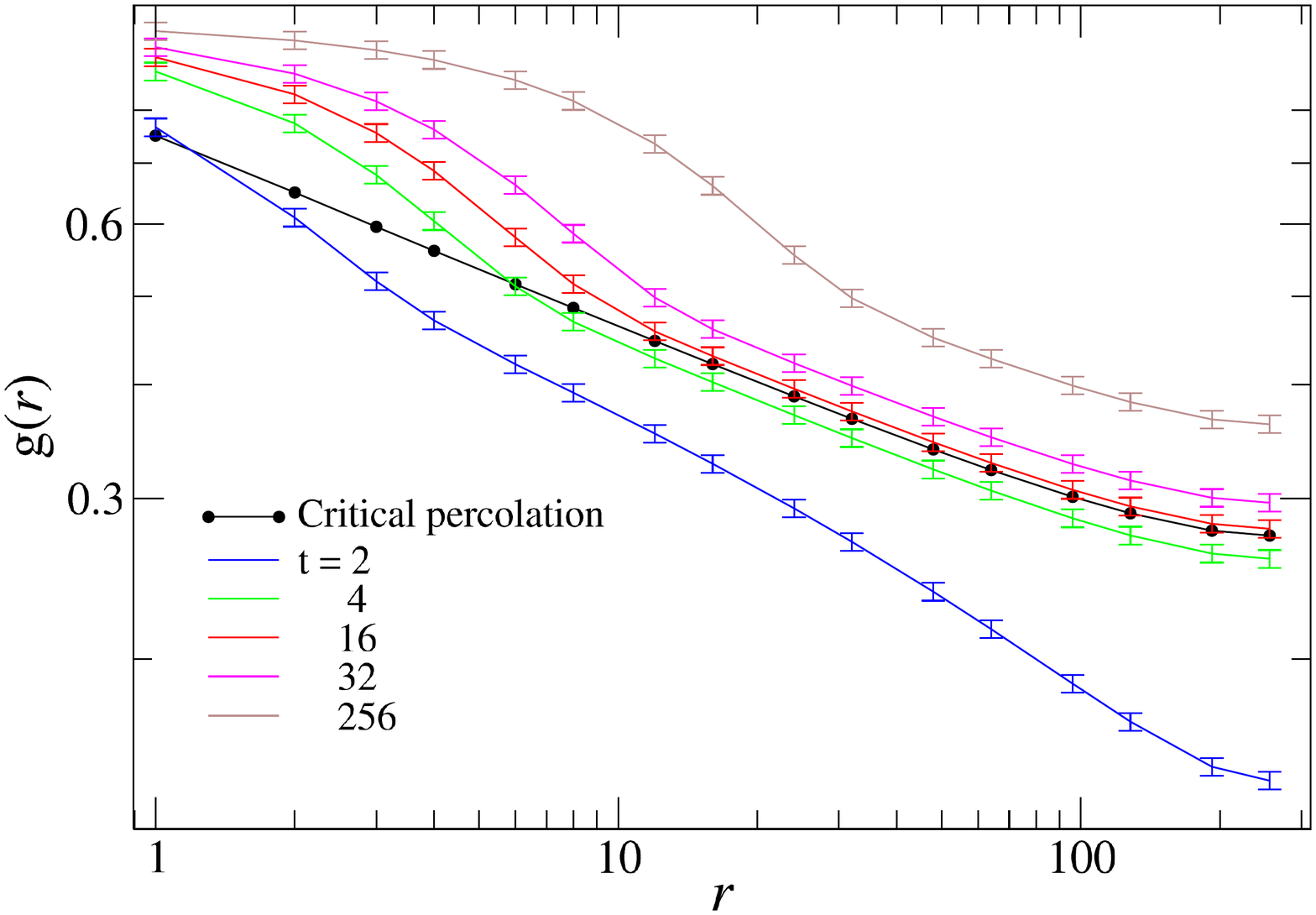}}}
\caption{Pair connectivity measured on a lattice at percolation 
threshold (black line) and for a coarsening homogeneous $2d$ IM (colored lines)  is plotted against the distance $r$. Measurement times are given in the key.
The lattice size is $L=512$ in both cases.}
\label{fcorr_pure}
\end{figure}

We find a similar behaviour for pair-connectedness, shown for the pure IM in Fig   \ref{fcorr_pure}.
We performed the measurement of $g(r)$ at different times on a system of size $L=512$.  For each curve except $t=2$, after an initial distance which increases with time, 
curves for coarsening have the same slope as that for percolation, the black line with slope $-2\Delta = - 10/48$ (at large distances the curves bends upward
as due to finite size effects).
Notice that this is not true for very small times, when the system is still very close to the initial condition.
The vertical shift of the curves for varying measurement times is interpreted in \cite{Corberi17} in the same way as for the winding angle.

In Figs. \ref{winding_pure_basic} and \ref{fcorr_pure}  we noticed a length, growing with time, after which the domains acquire the morphological properties of critical percolation.
It is natural to identify this length with the growing length $R(t)$, the typical size of the correlated regions (domains). Indeed, as mentioned in Sec. \ref{intro},  percolative properties can only be present a length scale over which the system is still uncorrelated. This interpretation can be formalised and quantitatively verified through scaling arguments \cite{Corberi17}.

\begin{figure}[!tbp]
  \centering
  \begin{minipage}[b]{0.48\textwidth}
   
    \includegraphics[width=\textwidth]{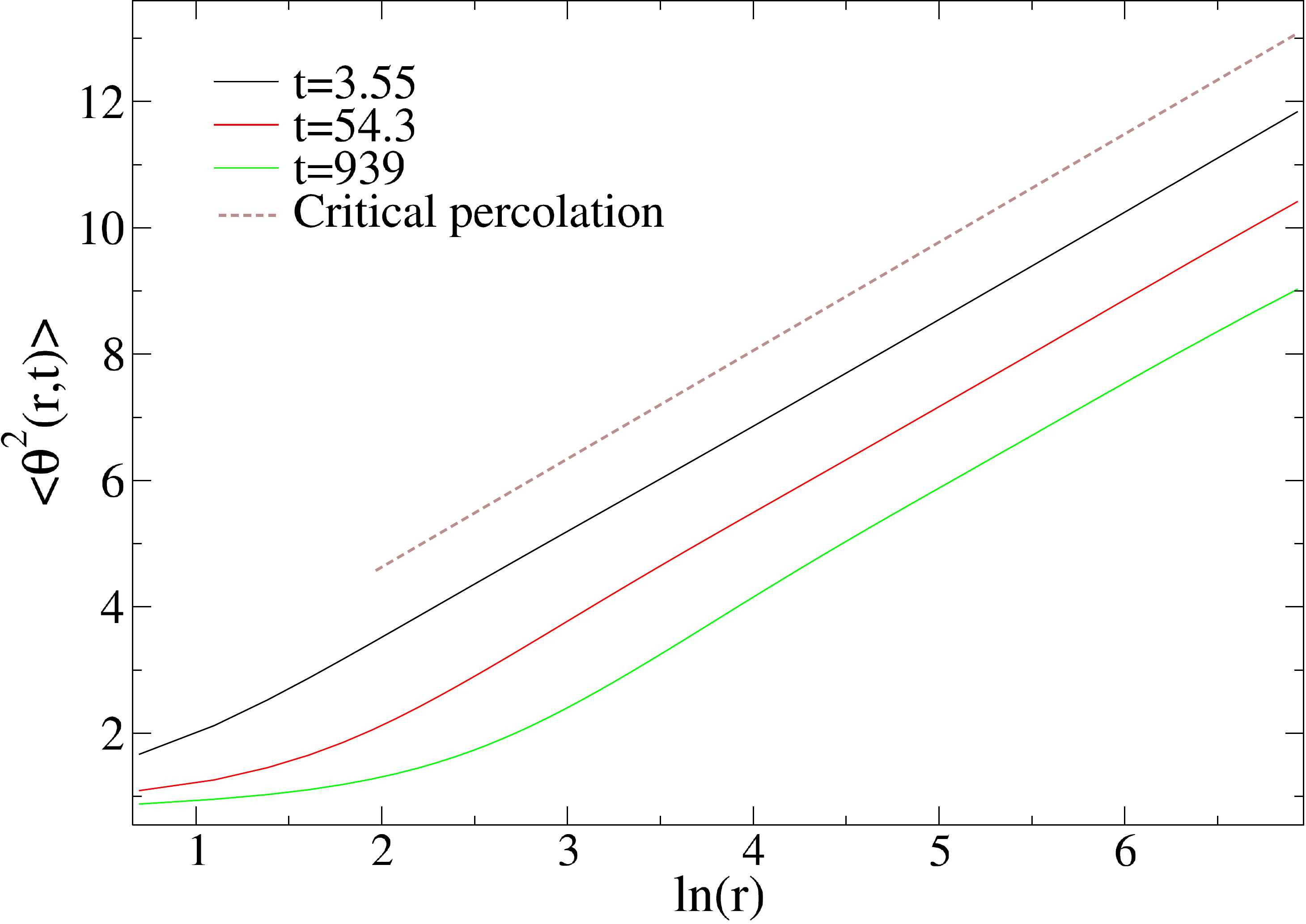}
    \caption{Same as Fig. \ref{winding_pure_basic}, but for RFIM with $\epsilon_{RF}=1$.}
     \label{winding_H1_basic}
  \end{minipage}
   \hspace{0.15cm}
\hfill
  \begin{minipage}[b]{0.495\textwidth}
    \includegraphics[width=\textwidth]{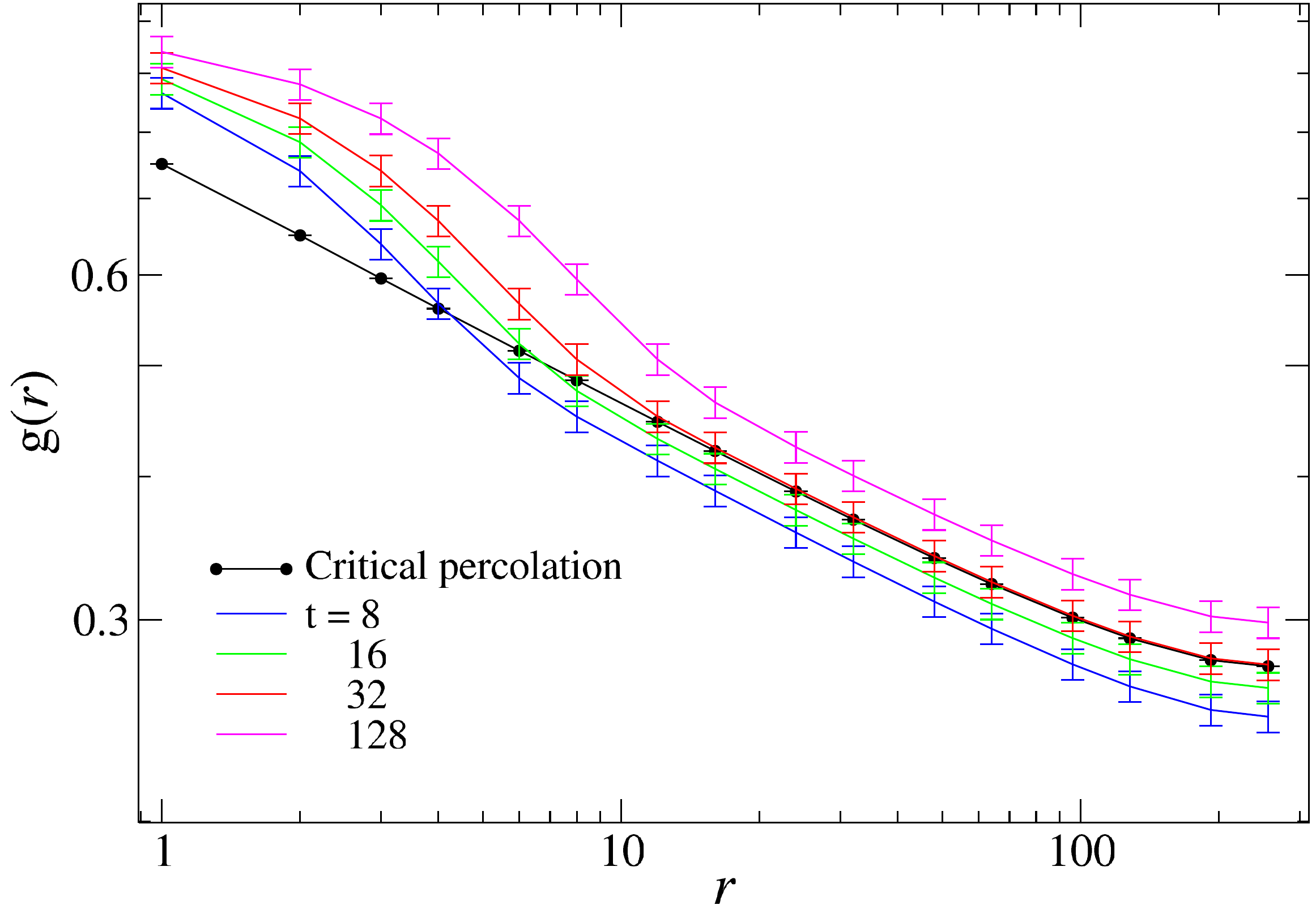}
    \caption{Same as Fig. \ref{fcorr_pure} but for a RFIM with $\epsilon_{RF}=1$.}
    \label{fcorr_H1}
  \end{minipage}
\end{figure}

Most importantly, we find analogous results for all the values of disorder strength considered, for both RFIM and RBIM. We only report data for the RFIM with $\epsilon=1$ in 
Figs. \ref{winding_H1_basic} and \ref{fcorr_H1}

We regard this as compelling evidence that the clusters arising in the coarsening process \textit{are} those of critical percolation, for both pure and disordered systems.
We can therefore conclude that percolation effects are quite robust, since they are not destroyed by the presence of the considered quenched disorder.
\section{Conclusions}
In this work we have investigated the relevance of percolative effects on the coarsening process of the $2d$ IM quenched from 
a completely disordered initial state to zero final temperature. We focused on the clean case and on two forms of quenched randomness, 
random bonds and random fields. We have used two quantities  -- winding angle and pair-connectedness -- which represent efficient tools to detect percolation effects.

In the light of our results, we can conclude that the influence of percolation on the coarsening dynamics of a quenched ferromagnet -- previously highlighted for homogeneous models -- also extends to systems where quenched disorder is present, systems that are much less understood. 
These results represent a first necessary step toward the generalisation of one of the few analytical theories of coarsening ~\cite{Sicilia08} -- currently available for pure systems -- that hinges on the percolative properties of the domains of the ordering system.

Finally, a natural extension of this work concerns randomly diluted models, a class of systems whose phase ordering kinetics 
presents subtleties and differences with respect to random field and random bond models \cite{Corberi13}.

\ack
F. I. and F. C.  thanks LPTHE Jussieu for hospitalty.
L. F. C.  is a member of Institut Universitaire de France, 
and she thanks the KITP University of California at Santa Barbara for hospitality. 
This research was  supported in part by the National Science Foundation under 
Grant No. PHY11-25915.

\end{document}